\documentclass[aip,graphicx]{revtex4-1}
\usepackage [latin1]{inputenc}
\draft 

\begin{document}

\title{New symmetry in higher curvature spacetimes} 

\author{Alcides Garat}
\affiliation{Former Professor at Universidad de la Rep\'{u}blica, Av. 18 de Julio 1824-1850, 11200 Montevideo, Uruguay.}

\date{\today}

\begin{abstract}
New symmetries have been found in Einstein-Maxwell spacetimes. New symmetries have also been found in imperfect fluid curved spacetimes. We will prove in this paper that we can extend these symmetries to spacetimes with higher curvature terms. Higher curvature theories are in many cases associated to dark energy for instance. We provide further justification for these higher curvature formulations through the existence of a new symmetry.
\end{abstract}




\maketitle 

\section{Introduction}
\label{intro}

New symmetries have been found for spacetimes where non-null electromagnetic fields and gravity exist. The origin of these new symmetries stem from the fact that the local group of electromagnetic gauge symmetries has been found to be isomorphic independently to two special local groups of tetrad transformations \cite{A,ROMP,SING,LomCon,ATGU,IWCP,AEO}. These local tetrad groups of transformations are realized inside two orthogonal planes at every point in spacetime. These new tetrad vectors are electromagnetic gauge dependent and when a local electromagnetic gauge transformation is implemented they transform inside these local orthogonal planes without leaving them thus ensuring the metric tensor invariance and the gravitational field explicit invariance. These orthogonal local planes defined by these tetrad vectors are the planes of diagonalization of the Einstein-Maxwell stress-energy tensor. All the vectors inside these unique local planes are eigenvectors of the stress-energy tensor. The plane spanned by the timelike and one spacelike vectors is plane one while the orthogonal plane is plane two. The local group of electromagnetic gauge transformations is isomorphic in the local plane one to the local group LB1 of local tetrad transformations. The group LB1 is made up by $SO(1,1) \times Z_{2} \times Z_{2}$ where $SO(1,1)$ is proper orthochronous. The first $Z_{2}$ is given by $\{I_{2 \times 2}, -I_{2 \times 2}\}$ and the second $Z_{2}$ is given by $\{I_{2 \times 2}, \mbox{the swap}\: (01|10)\}$. There are two discrete transformations. One of them that we designated as $-I_{2 \times 2}$ is the full inversion two by two and the other a reflection designated as $\mbox{the swap}\: (01|10)$ and given by $\Lambda^{o}_{\:\:o} = 0$, $\Lambda^{o}_{\:\:1} = 1$, $\Lambda^{1}_{\:\:o} = 1$,  $\Lambda^{1}_{\:\:1} = 0$ which is not a Lorentz transformation because it is a reflection. We would have to add in order to complete the image of the map $SO(1,1) \times Z_{2} \times Z_{2}\: \bigoplus \: \{light\:cone\:gauge\}$ where the light cone gauge includes the inhomogeneous two solutions to the differential equations in the local future and past light cones established in reference \cite{SING} where the reflection through the asymptote $cT=X$ on plane one will produce two more identical inhomogeneous solutions. A total of four. In the local orthogonal plane two and independently from the mapping in the plane one, the local group of electromagnetic gauge transformations is mapped onto the local group of spatial tetrad rotations LB2 which is just $SO(2)$. LB stands for Lorentz blade or plane. We can also consider spacetimes with imperfect fluids with vorticity where we also study the four-velocity gauge-like transformations that leave the vorticity tensor invariant exactly as in the electromagnetic case but replacing the electromagnetic potential for the fluid four-velocity. If we also consider appropriate local transformations in the energy-density, heat flux, viscous stresses we can prove that the Einstein equations are invariant under this symmetry. We will use both the metric tensor invariance in Einstein-Maxwell spacetimes as well as the invariance of the metric tensor under four-velocity gauge-like groups of transformations for imperfect fluids in order to prove that the previously found symmetries are also symmetries in higher curvature spacetimes with the same stress-energy tensors.

\subsection{Einstein-Maxwell spacetimes}
\label{einsteinmaxwell}

The metric tensors are invariant under LB1 and LB2 in these general Einstein-Maxwell four-dimensional curved Lorentz spacetimes where the only observation is that the electromagnetic fields are non-null as proven in detail in references \cite{A,ROMP,SING,ATGU,IWCP,LomCon,AEO}. We use a signature $-+++$ and if $F_{\mu\nu}$ is the electromagnetic field then $f_{\mu\nu}= (G^{1/2} / c^2) \: F_{\mu\nu}$ is the geometrized electromagnetic field. If the metric tensors in Einstein-Maxwell spacetimes where electromagnetic fields are non-null are invariant under these symmetries, then the covariant derivatives are also invariant and the Riemann and Ricci tensors are also invariant. In standard Einstein-Maxwell spacetimes this result means that the left hand side of the differential equations is manifestly invariant under this local group of electromagnetic gauge transformations. The right hand side, that is the Einstein-Maxwell stress-energy tensor is locally invariant under the local group of electromagnetic gauge transformations because the electromagnetic field and the metric tensors are also invariant. It is therefore automatic to see that local invariance under this same local group will remain if we consider modified gravity theories like Lagrangians including terms like $R + a_{2}\:R^{2} + a_{3}\:R^{3}$ with $a_{2}$ and $a_{3}$ constants or Lagrangians including terms like
$a\:R_{\mu\nu}\:R^{\mu\nu}+b\:R_{\mu\nu\gamma\delta}\:R^{\mu\nu\gamma\delta}$ with $a$ and $b$ also constants. Theories also known as $f(R,T)$ include higher curvature terms \cite{CSEN,SDSY,TDT,Myers,SMCR,MTT,TMRS,TM,ZC,FC1,FC2,FC3} and the same conclusions will ensue. In this theories we just highlight that the trace of the Einstein-Maxwell stress-energy tensor will be zero. The references \cite{CSEN,SDSY,TDT,Myers,SMCR,MTT,TMRS,TM,ZC,FC1,FC2,FC3} are just a few examples in a huge literature library.

\subsection{Imperfect fluid spacetimes}
\label{imperfectfluid}

We will consider next an imperfect fluid stress-energy tensor of the kind \cite{JO},

\begin{equation}
T^{imp}_{\mu\nu}= (\rho + p)\:u_{\mu}\:u_{\nu} + p\:g_{\mu\nu} + (q_{\mu}\:u_{\nu} + q_{\nu}\:u_{\mu}) + \tau_{\mu\nu} \ , \label{SETIMP}
\end{equation}

\noindent where $q_{\mu}$ is the heat flux relative to the fluid four-velocity $u_{\mu}$ and the viscous stress-energy tensor $\tau_{\mu\nu}$ is provided by \cite{JO},

\begin{equation}
\tau_{\mu\nu}= -\eta\:\left(u_{\mu;\nu} + u_{\nu;\mu} + u_{\mu}\:u^{\alpha}\:u_{\nu;\alpha} + u_{\nu}\:u^{\alpha}\:u_{\mu;\alpha}\right) - (\zeta-\frac{2}{3}\:\eta)\:u^{\alpha}_{\:\:;\alpha}\:(g_{\mu\nu} + u_{\mu}\:u_{\nu}) \ .\label{SETIMPVISC}
\end{equation}

\noindent The parameter $\eta$ is the coefficient of shear viscosity and the parameter $\zeta$ is the coefficient of bulk viscosity. Under the four-velocity gauge-like transformation $u^{\alpha} \rightarrow u^{\alpha}+\Lambda_{,\beta}\:g^{\beta\alpha}$ (we will use the notation $\Lambda^{\alpha}=\Lambda_{,\beta}\:g^{\beta\alpha}$ where $\Lambda$ is a local scalar) the stress-energy tensor (\ref{SETIMP}) will change and will not be invariant if this is the only transformation that we implement. Let us remember that the Einstein-Imperfect Fluid equations are given now by equation,

\begin{eqnarray}
R_{\mu\nu} - \frac{1}{2}\:g_{\mu\nu}\:R &=& T^{imp}_{\mu\nu} \ . \label{EFEIMP}
\end{eqnarray}

\noindent We have already found that the local tetrads built for the imperfect fluid with vorticity \cite{EG,AC,F,CD,GQW,MSS} under local four-velocity gauge-like transformations $u^{\alpha} \rightarrow u^{\alpha}+\Lambda^{\alpha}$ leave the metric tensor invariant and this result implies that the left hand side of equation (\ref{EFEIMP}) is locally invariant under this group of Abelian transformations. We have also demonstrated how that is possible and shown explicitly how the energy-density, the pressure, the heat flux and the viscous stresses transform simultaneously in order to achieve invariance on the right hand side. When we carry out the transformations $u^{\alpha} \rightarrow u^{\alpha}+\Lambda^{\alpha}$ on the right hand side of equation (\ref{EFEIMP}) we notice that the tensor $T^{imp}_{\mu\nu}$ is not invariant if this is the only kind of transformation that we implement. In order to make it locally invariant we require the following additional set of transformations,

\begin{eqnarray}
\rho &\rightarrow& \rho + \widetilde{\rho} \label{rhotransf}\\
p &\rightarrow& p + \widetilde{p} \label{ptransf}\\
q^{\mu} &\rightarrow& q^{\mu} + \widetilde{q}^{\mu} \label{heattransf}\\
\tau^{\mu\nu}(u) &\rightarrow& \tau^{\mu\nu}(u+\Lambda) + \widetilde{\tau}^{\mu\nu} \ . \label{viscoustransf}
\end{eqnarray}

\noindent By $\tau^{\mu\nu}(u)$ we mean exactly the expression (\ref{SETIMPVISC}) while $\tau^{\mu\nu}(u+\Lambda)$ means equation (\ref{SETIMPVISC}) under the transformation $u^{\alpha} \rightarrow u^{\alpha}+\Lambda^{\alpha}$. We have proposed notation that shortens the explicit writing of long expressions with many terms even though clear in its content. We also know from the outset of this formulation that there is an equation of state $p(\rho)$ and there is also another equation of state $\widetilde{p}(\widetilde{\rho})$. When invariance is imposed on the stress-energy tensor $T^{imp}_{\mu\nu}$ we obtain ten equations for the fifteen variables $\widetilde{\rho}$, $\widetilde{q}^{\mu}$ and $\widetilde{\tau}^{\mu\nu}$. In addition we also know that there are five more equations $q^{\alpha}\:u_{\alpha}=0$ and $\tau^{\mu\nu}\:u_{\nu}=0$. When we apply these last two equations to the case under $u^{\alpha} \rightarrow u^{\alpha}+\Lambda^{\alpha}$ we find $(q^{\mu} + \widetilde{q}^{\mu})\:(u_{\mu}+\Lambda_{\mu})=0$ and $(\tau^{\mu\nu}(u+\Lambda) + \widetilde{\tau}^{\mu\nu})\:(u_{\mu}+\Lambda_{\mu})=0$. By employing these last two ``transverse'' equations plus the original ones $q^{\alpha}\:u_{\alpha}=0$ and $\tau^{\mu\nu}\:u_{\nu}=0$ we obtain,

\begin{eqnarray}
q^{\mu}\:\Lambda_{\mu} + \widetilde{q}^{\mu}\:(u_{\mu}+\Lambda_{\mu})&=&0 \label{conditionsheat}\\
\tau^{\mu\nu}(u)\:\Lambda_{\nu} + \tau^{\mu\nu}(\Lambda)\:(u_{\nu}+\Lambda_{\nu}) + \widetilde{\tau}^{\mu\nu}\:(u_{\nu}+\Lambda_{\nu})&=&0 \ . \label{conditionsvisc}
\end{eqnarray}

\noindent The notation $\tau^{\mu\nu}(\Lambda)$ represent all the terms of the style $\Lambda_{\mu}\:u_{\nu;\alpha}\:u^{\alpha}+\Lambda_{\nu}\:u_{\mu;\alpha}\:u^{\alpha}$ or $u_{\mu}\:\Lambda_{\nu;\alpha}\:\Lambda^{\alpha}+u_{\nu}\:\Lambda_{\mu;\alpha}\:\Lambda^{\alpha}$ or $\Lambda_{\mu}\:\Lambda_{\nu;\alpha}\:\Lambda^{\alpha}+\Lambda_{\nu}\:\Lambda_{\mu;\alpha}\:\Lambda^{\alpha}$ just showing a few examples. The tensor $\tau^{\mu\nu}(u)$ represents (\ref{SETIMPVISC}) and we can write $\tau^{\mu\nu}(u+\Lambda)=\tau^{\mu\nu}(u)+\tau^{\mu\nu}(\Lambda)$.

In reference \cite{AGNS} the explicit expressions for $\widetilde{\rho}$, $\widetilde{p}$, $\widetilde{q}^{\mu}$ and $\widetilde{\tau}^{\mu\nu}$ have been found. Additionally and based on this principle of symmetry we have also paid special attention to the construction of a vorticity stress-energy tensor which is also invariant by construction under this same group of local four-velocity gauge-like transformations. An application on neutron stars was developed in order to show the simplifications brought about by these new tetrads \cite{AGNS,ACD,ENV}. Because and we quote from reference \cite{SCNO} ``The macroscopic neutron vorticity $\varpi_{n}$ (we use Greek letters for spacetime indices) $\varpi_{n}=\sqrt{\frac{\varpi_{\mu\nu}\:\varpi^{\mu\nu}}{2}}$ where the vorticity 2-form $\varpi_{n}$ is defined by $\varpi_{\mu\nu}=\nabla_{\mu}p_{\nu}^{n}-\nabla_{\nu}p_{\mu}^{n}$, $p_{\mu}^{n}$ denoting the conjugate superfluid momentum. We note here that, on length scales smaller than the intervortex separation $d_{v}$, typically of the order of $d_{v} \sim n_{v}^{-1/2} \sim 10^{-3} cm$ (see Eq. (1)), $\varpi_{n}$ strictly vanishes because $p_{\mu}^{n}$ should be locally proportional to the gradient of a quantum scalar phase. Nevertheless, on the large scales we are interested in here, the neutron vorticity 2-form is non-vanishing, as well as its corresponding scalar amplitude $\varpi_{n}$'', see references \cite{SCNO,CH,LP,HM,BC1,LSC,CPG,EGAPJ,ASC,SON,CS,CC,PA}. We can state the following conclusions as a summary.

\begin{itemize}

\item New orthonormal tetrad for perfect fluids in four-dimensional curved Lorentzian spacetimes. This tetrad diagonalizes locally and covariantly the perfect fluid stress-energy tensor. Astrophysical applications in spacetime evolution \cite{AEO} and \cite{AGNS,ACD,ENV}.

\item New symmetry group associated with four-velocity gauge-like local transformations. In order to achieve this goal of finding this new local symmetry, new local transformations are found for the energy-density, pressure, heat flux and viscous stresses.

\item Maximum simplification of relevant tensors and field equations, see reference \cite{AGNS}.

\item New tetrads encode gravitational and fluid gauge information.

\item We are introducing an explicit ``link'' between the ``fluid four-velocity gauge-like'' groups of transformations and the ``spacetime'' groups of transformations exactly as in Einstein-Maxwell spacetimes \cite{A,ROMP,SING,LomCon,ATGU}.

\item New vorticity stress-energy tensor proposed on the basis of the new local symmetry \cite{AGNS}.

\item Applications to neutron star evolution in relativistic astrophysics  \cite{AGNS}.

\item  Higher curvature theories are in many cases associated to dark energy for instance. We provide further justification for these higher curvature formulations through the existence of a new symmetry.

\item Experiments will be proposed on the basis of fluid vorticity and four-velocity in order to change the local causality nature of spacetime analogously to manuscripts \cite{SCR,FULL} for the electromagnetic case.

\end{itemize}

We can thus summarize the results in previous references \cite{AGNS,ACD,ENV} where we have proved the following results,

\newtheorem {guesslb1} {Theorem}
\newtheorem {guesslb2}[guesslb1] {Theorem}
\begin{guesslb1}
The mapping between the local group of four-velocity gauge-like transformations and the local group LB1 defined above is isomorphic.
\end{guesslb1}

\begin{guesslb2}
The mapping between the local group of four-velocity gauge-like transformations and the local group LB2 defined above is isomorphic.
\end{guesslb2}

\begin{guesslb2}
The imperfect fluid stress-energy tensor and independently the vorticity stress-energy tensor are invariant under the group of four-velocity gauge-like transformations as long as we also implement the local transformations to energy-density, pressure, heat flux and viscous stresses presented in equations (\ref{rhotransf}-\ref{viscoustransf}).
\end{guesslb2}

When we put together all of these results we readily notice that if we consider modified gravity theories as in section \ref{einsteinmaxwell} it is again automatic to see that local invariance under this same local group of four-velocity gauge-like transformations will remain for the modified Einstein-Modified-Fluid equations as well because the metric tensor will also be locally invariant.

\section{Conclusions}
\label{conclusions}

We have managed to prove that the local symmetries already found in Einstein-Maxwell and Einstein-Imperfect Fluid spacetimes remain as local symmetries also in the case of modified gravity theories. This is encouraging in several fields of research. For instance in cosmology, relativistic astrophysics, systems involving electromagnetic fields, etc. Higher curvature theories are in many cases associated to dark energy for instance. We provide further justification for these higher curvature formulations through the existence of a new symmetry. These symmetries also justify the existence of a vorticity tensor in Einstein-Imperfect Fluid spacetimes. It is remarkable that these local symmetries remain also for the whole variety of modified gravities and the profound reason is that these local groups of Abelian symmetries are manifest local symmetries of the metric tensor and of the gravitational field. We can justify this way the employment of modified gravities with the additional notion of symmetries and only experiments and measurements will distinguish the true formulation for relativistic systems.

\section{Declaration of interest statement}
\label{interest}

The authors declare that they have no known competing financial interests or personal relationships that could have appeared to influence the work reported in this paper.

\section{Data availability statement}
\label{data}

There is no data to be reported in this paper.


\end{document}